\documentclass[a4paper]{PoS}

\usepackage{todonotes}
\usepackage{textcomp}

\title{Calibration of the absolute amplitude scale of the Tunka Radio Extension (ICRC 2015)}

\ShortTitle{Tunka-Rex Calibration}

\author{
\speaker{R.~Hiller}$^{1}$, P.A.~Bezyazeekov$^{2}$, N.M.~Budnev$^{2}$, O.A.~Gress$^{2}$, A.~Haungs$^{1}$, T.~Huege$^{1}$, Y.~Kazarina$^{2}$, M.~Kleifges$^{3}$, E.N.~Konstantinov$^{2}$, E.E.~Korosteleva$^{4}$, D.~Kostunin$^{1}$, O.~Kr\"omer$^{3}$, L.A.~Kuzmichev$^{4}$, N.~Lubsandorzhiev$^{4}$, R.R.~Mirgazov$^{2}$, R.~Monkhoev$^{2}$, A.~Pakhorukov$^{2}$, L.~Pankov$^{2}$, V.V.~Prosin$^{4}$, G.I.~Rubtsov$^{5}$, F.G.~Schr\"oder$^{1}$, R.~Wischnewski$^{6}$, A.~Zagorodnikov$^{2}$
- 
Tunka-Rex Collaboration \\
\llap{$^1$} Institut f\"ur Kernphysik, Karlsruhe Institute of Technology (KIT), Germany\\
\llap{$^2$} Institute of Applied Physics ISU, Irkutsk, Russia\\
\llap{$^3$} Institut f\"ur Prozessdatenverarbeitung und Elektronik, Karlsruhe Inst.~of Tech.~(KIT), Germany\\
\llap{$^4$} Skobeltsyn Institute of Nuclear Physics MSU, Moscow, Russia\\
\llap{$^5$} Institute for Nuclear Research of the Russian Academy of Sciences, Moscow, Russia\\
\llap{$^6$} DESY, Zeuthen, Germany\\
E-mail: \email{roman.hiller@kit.edu}       
}

\abstract{
The Tunka Radio Extension (Tunka-Rex) is an array of 44 radio antenna stations, distributed over 3\,km$^{2}$, constituting a radio detector for air showers with an energy threshold around 10$^{17}$\,eV. 
It is an extension to Tunka-133, an air-Cherenkov detector in Siberia, which is used as an external trigger for Tunka-Rex and 
provides a reliable reconstruction of energy and shower maximum.
Each antenna station consists of two perpendicularly aligned active antennas, called SALLAs. 
An antenna calibration of the SALLA with a commercial reference source enables us to 
reconstruct the detected radio signal on an absolute scale.
Since the same reference source was used for the calibration of LOPES and, in a calibration campaign in 2014, also for LOFAR, 
these three experiments now have a consistent calibration and, therefore, absolute scale.
This was a key ingredient to resolve a longer standing contradiction between measurements of two calibrated experiments.
We will present how the calibration was performed and compare radio measurements of air showers from Tunka-Rex to model calculations 
with the radio simulation code CoREAS, confirming it within the scale uncertainty of the calibration of 18\%.
}

\FullConference{The 34th International Cosmic Ray Conference,\\
		30 July- 6 August, 2015\\
		The Hague, The Netherlands}

\begin{document}

\section{Introduction}
To overcome restrictions of commonly used detectors for cosmic ray air showers, 
new detector types and detection principles are being investigated since the inception of air shower detection.
One promising candidate for a next generation air shower detector, or part of a hybrid detector, is the radio detection technique. 
It is based on the radio emission of air showers: 
Mainly due to the geomagnetic deflection of charged particles, a short electromagnetic pulse in the radio frequency range is emitted~\cite{geomag}.
A small, but still significant contribution to the pulse from the Askaryan effect~\cite{askaryan}, originating from a varying net charge, was experimentally confirmed lately~\cite{codalemaaskaryan, lofaraskaryan, aeraaskaryan}.

The resulting pulses can be detected with an antenna array.
The main advantages of this method are its almost full duty cycle and the indicated high sensitivity to air shower parameters. 
Because of the high transparency of the atmosphere for radio frequencies, 
it is also possibly well suited for high observation depths or highly inclined showers.

The first generation of digital radio arrays~\cite{codalema, lopes} proved their general feasibility and sensitivity to air-shower parameters, 
while also fueling progress in the theoretical understanding of the radio emission.
Contemporary radio arrays now have to either show their competitiveness to established detection principles or refine the methods and technology to reach it.

The Tunka Radio Extension (Tunka-Rex) is a digital radio-antenna array in the Tunka valley in Siberia, Russia.
It is co-located with Tunka-133~\cite{t133}, a photomultiplier array for the detection of cosmic-ray air showers via their Cherenkov emission in air.
The detector concept of Tunka-Rex is to record the radio signal of a shower in coincidence with the air-Cherenkov signal, 
by triggering the antenna array with Tunka-133. 
Thereby obtaining a reliable trigger, as well as the shower parameters from an independent measurement with known precision to compare
the radio reconstruction to. 
However, Tunka-133 is limited in its operation time to moonless nights during winter.
To overcome this limitation, in near future, Tunka-Rex will also be triggered by Tunka-Grande~\cite{taiga}, 
a recently installed scintillator-based detector for cosmic-ray air showers at the site, capable of operation around-the-clock.

One goal of Tunka-Rex is to obtain a calibrated measurement of the radio signal.
This enables further testing and improvement of the modeling of the radio signal, 
as well as determining its absolute scale for comparison to other radio experiments.
For this purpose we performed an antenna calibration with a reference source.

In this paper we describe the Tunka-Rex antenna station and its calibration.
Finally we compare measured radio amplitudes from air-shower events to simulated ones with the simulation software CoREAS.
\section{Detector setup}
	\begin{figure}[tb]
		\center
		\includegraphics[width=0.7\textwidth]{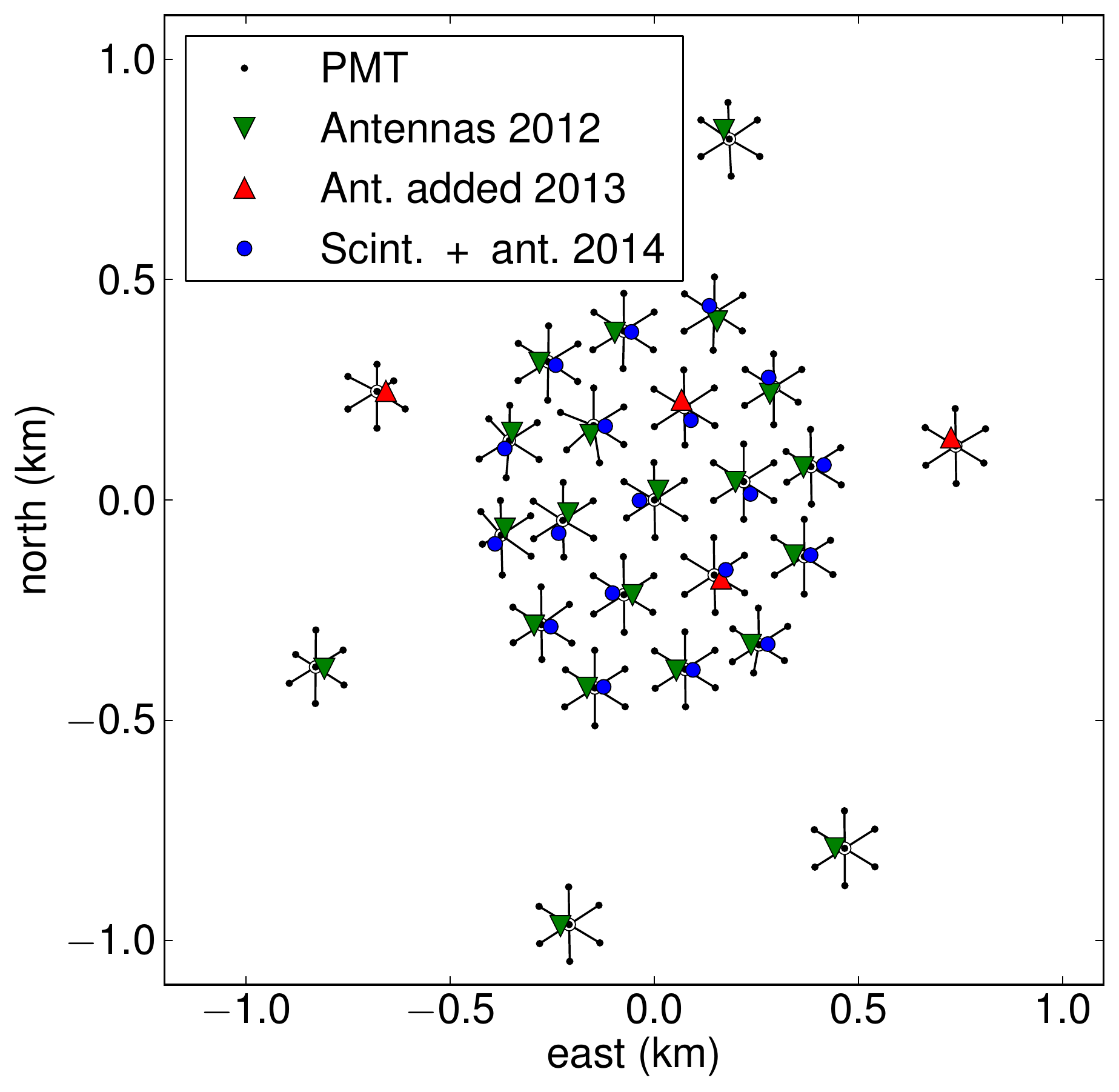}
		\caption{The layout of Tunka-133 and Tunka-Rex in its different stages.
		Tunka-133 is organized in hexagonal clusters. 
		Each cluster is equipped with one Tunka-Rex antenna station and 
		a scintillator station with another antenna station connected to it.}
		\label{fig:layout}
	\end{figure}
The layout of Tunka-Rex is depicted in Fig.\ref{fig:layout}. Its central part consists of 19 antenna stations, 
distributed over 1\,km$^{2}$. These antennas are currently triggered by Tunka-133.
There is one antenna connected to each of the 19 clusters of Tunka-133.
A cluster consists of a group of 7 PMTs in a hexagonal pattern with common housing for electronics and data acquisition (DAQ) in its center.
Together with 6 more distant antenna stations Tunka-Rex covers 3\,km$^{2}$, but these outer antennas are only relevant for the highest energies,
since the antenna spacing decreases from roughly 200\,m to 500\,m in this outer region.
With the Tunka-Grande scintillator extension each cluster was extended by a scintillator station with another antenna station, totaling 
44 antennas, for the whole array. Once Tunka-Grande is fully comissioned, it will trigger the whole array.

Each Tunka antenna station consists of two channels, each with a SALLA~\cite{AERAantenna} antenna, aligned perpendicularly to each other.
By reconstructing the incoming direction of the radio signal from the relative arrival time, 
the two channels enable the reconstruction of the electric field vector.

\begin{figure}[tb]
	\center
	\includegraphics[width=0.7\textwidth]{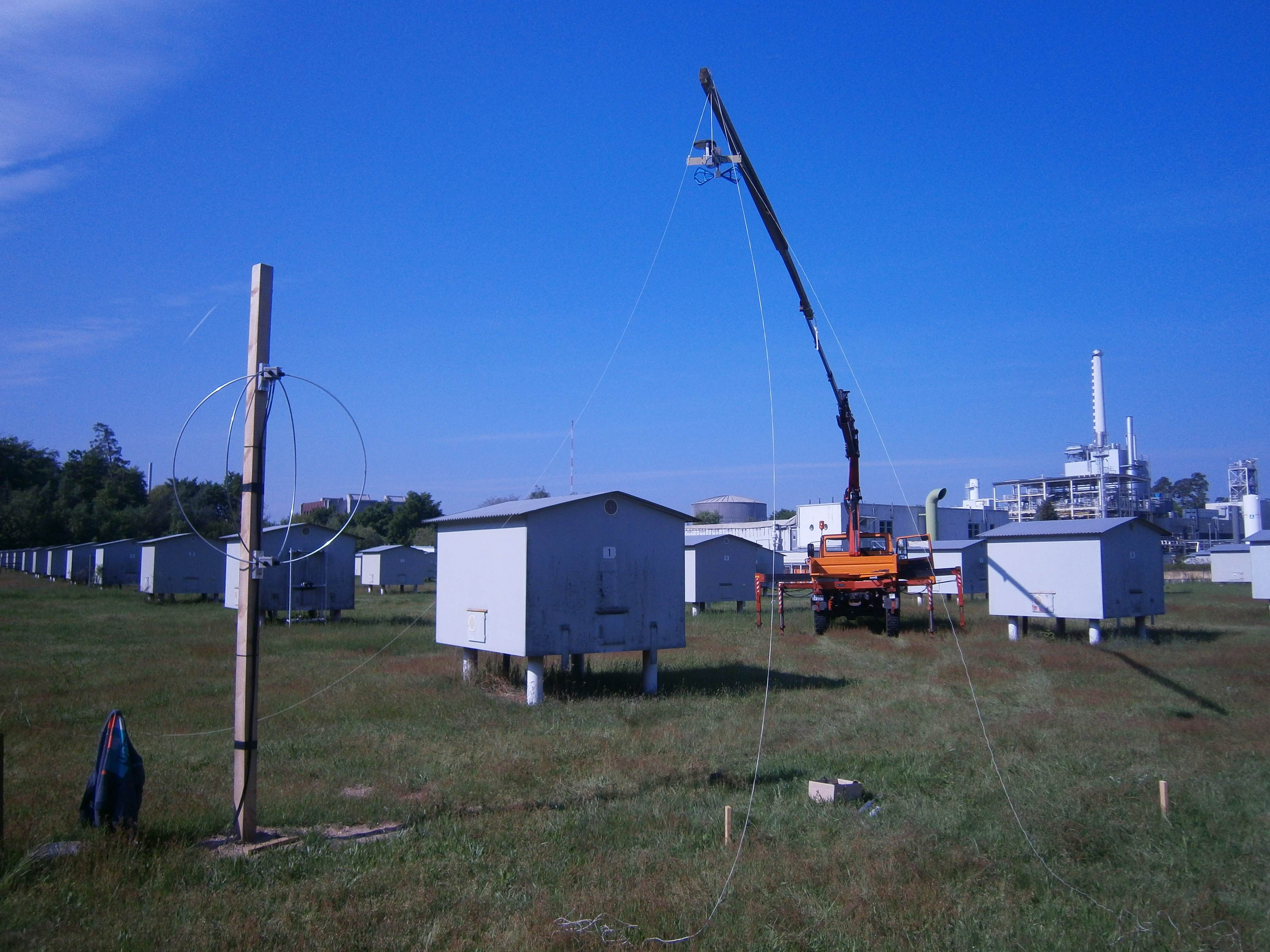}
	\caption{Setup of the Tunka-Rex antenna calibration in Karlsruhe. A crane holds a reference source above the antenna
	aligned from the ground using strings. With the known emitted field strength of the reference source, 
	the characteristics of the SALLA can be determined.}
	\label{fig:salla}
\end{figure}

Each SALLA is directly connected to a low-noise-amplifier (LNA), in the upper antenna box 
(see Fig. \ref{fig:salla}), enhancing the signal by about 22\,dB.
From there the signal is transferred via coaxial cable to the cluster center in 20\,m distance.
The cluster center is heated to ensure stable operation for the contained electronics in cold winter nights. 
There, the signal is filtered to the frequency range 35-76\,MHz and enhanced by another 32\,dB by a filter-amplifier.
Finally, the so far analog signal is digitized at 200\,Ms/s with 12 bit depth, and 
a $\approx 5$\,\textmu s trace containing 1024 samples is recorded around the trigger time of Tunka-133.
\section{Antenna calibration}
The determine the electric field on an absolute scale from the ADC counts of the trace, 
the impact of the full hardware chain has to be understood and inverted.
From a formal point of view, the impact of a linear system on an incoming signal 
is described by the convolution of the incoming signal with its response function.
For an antenna, this response function depends on the incoming direction.
For dipole like antennas, the vector effective length $\vec H$ is commonly used as response function 
and especially convenient if the polarization of the signal is of interest.
In case of an antenna the incoming signal is an electric field $\vec E$ and the outgoing signal is a voltage $V$.
In the frequency domain the convolution becomes a simple point-wise multiplication
\begin{equation}
V(\nu)=\vec H(\nu)\cdot \vec E(\nu).
\label{eq:calib_simple}
\end{equation}
Provided two measurements of $V$ from the two perpendicular antenna channels with linearly independent sensitivity to the polarization,
this equation can be inverted to reconstruct all three components of $\vec E$, if the incoming direction and $\vec H$ is known.
While the incoming direction in each event is determined from the arrival times, assuming the incoming direction of the shower and the radio signal coincide, 
$\vec H$ remains the same for all events and has to be determined only once.

To obtain $\vec H$, we performed a calibration of the full analog hardware chain, 
including the antenna, with a reference source.
Therefore, a complete Tunka-Rex antenna station was deployed at the Karlsruhe Institute of Technology, Germany.
On the Tunka site, the station is plugged into the existing DAQ, which was not available in Karlsruhe, 
so we had to rely on a oscilloscope for data acquisition.
This makes the calibration almost end-to-end, except for the DAQ, containing the analog-to-digital converter, which was calibrated independently.

The reference source we used was the same as the one used for the calibration of LOPES~\cite{lopescalib} and recently LOFAR~\cite{lofarcalib}.
The available calibration of the reference source was performed under free-field conditions, with a high impact from ground reflections.
Therefore, we obtained new calibration values for the reference source for free-space conditions, 
which are also used for an update of the LOPES results~\cite{lopescalibICRC}.
Due to the use of the same reference source, these three experiments now have the same amplitude scale 
and can be compared neglecting the scale uncertainty due to the calibration of the reference source.

The reference source is the commercially available biconical antenna and 
signal generator VSQ1000+DPA4000 by Schaffner Electrotest GmbH (now Teseq).
It produces a signal linearly polarized along the antenna.
In the time domain the signal is a comb with one puls per \textmu s.
In the frequency domain this results in a comb, consisting of peaks with 1\,MHz spacing. 
The absolute height of the peaks, i.e. the amplitudes at $r_{ref}=10$\,m for the respective frequencies are given by the manufacturer with an accuracy of 16\% (one sigma), making up for the dominant uncertainty in the amplitude reconstruction.

For the calibration, the source is placed at different zenith angles above the antenna in >10\,m 
distance to the antenna by a crane, 
standing in 20\,m distance, holding the source with a wooden spacer.
With a differential GPS the distance $r$ between the reference source and the SALLA are determined with a precision of several 10\,cm.
The source is aligned via strings from ground along the polarization axis of maximum sensitivity
of one of the SALLAs.
So the one channel receives maximal power, while the other one receives minimal power.
We can obtain $\vec H(\nu)$ via:
\begin{equation}
\vec H(\nu)=\frac{V(\nu)}{\left|\vec E(\nu)\right|}\cdot \frac{r}{r_{ref}}\cdot \vec p.
\label{eq:calib}
\end{equation}
$\left|\vec E(\nu)\right|$ is the absolute value of the electric field strength of the reference source.
To obtain the polarization axis of maximum sensitivity $\vec p$, 
an antenna simulation of the passive antenna structure is performed with the NEC2 code. 
It turns out, that the sensitivity of the SALLA closely follows the
pattern of a horizontal dipole in the antenna arc plane.

Measuring the response of the analog hardware separately with a network analyzer~\cite{trexHWICRC13},
we can subtract it from the total response to extract the VEL of the bare antenna.
\begin{figure}[tb]
	\center
	\includegraphics[width=0.7\textwidth]{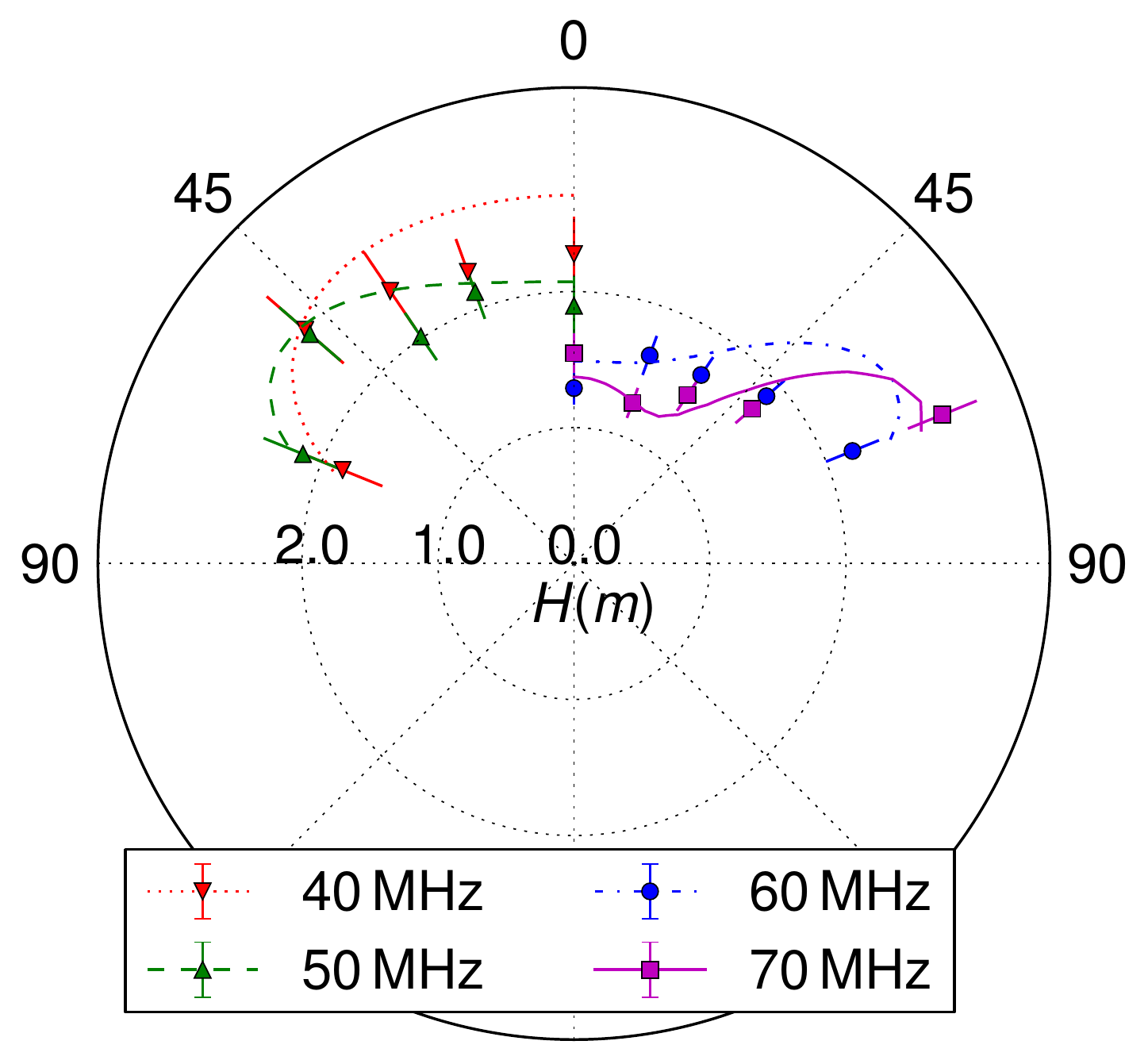}
	\caption{Measured vector effective lengths of the bare SALLA (markers) and simulated ones (lines).
	The left-right seperation is only for illustration, since all curves are mirror-symmetric.}
	\label{fig:NEC}
\end{figure}
In Fig.~\ref{fig:NEC}, the simulated VEL is shown against the measured one.
The relative deviation of measured from simulated VEL over different zenith angles is 12\%, averaged over all relevant frequencies, used as an estimated for the precision of the directivity model.
The scale deviation is 6\%, which is in the expected order due to the calibration uncertainty of the reference source of 16\%.

For the calibration we normalize the simulated antenna pattern frequency-wise 
to the mean of all calibration measurements at different zenith angles.
The total scale uncertainty is 18\%, resulting from the squared sums of
16\% from the uncertainty of the calibration of the reference source,
6\% from the temperature during the calibration and 4\% from positioning and alignment of the reference source.
The uncertainty on the absolute amplitude reconstruction on antenna level is 22\%. Besides the 18\% on the scale, it includes 12\% from the antenna model and some smaller contributions from temperature, 
antenna alignment and uncertainties in the production and deployment of the antenna stations (details in \cite{TRexcalib}).

\section{Amplitude comparison}
After the calibration we compare model calculations of the radio emission to the actual measured amplitudes.
Therefore, we take the Tunka-133 data set of the first season, 2012/2013, 
where the reconstructed primary energy is 
above 10$^{16}$\,eV and an zenith angle below 50$^{\circ}$.

We then cut the frequency range to the realized bandwidth of 35-76\,MHz
and remove some single frequency entries from the spectrum at every 5\,MHz, 
where we frequently observe interferences.
Then the full hardware response is inverted.
A noise level is determined by the RMS of a 500\,ns window before the signal window
and a signal $S$ by the peak of the Hilbert envelope in the signal window.
As radio event only those qualify which have at least 3 stations with a signal-to-noise ratio ($S^2/N^2$)
of at least 10, which enables the reconstruction of the shower axis from the relative arrival times of the pulses.
The shower axis then has to agree with the on reconstructed by Tunka-133 within 5$^{\circ}$.

This leaves 92 events in the data set for which we perform simulations of the radio signal with CoREAS~\cite{CoREAS}.
The events are simulated for proton and iron primaries with initial parameters from the Tunka-133 reconstruction.
Since the depth of the shower maximum cannot be directly set in the simulation, but emerges from other parameters and shower fluctuations,
we repeat the simulations several times and take only simulations into account, which happen to have their shower maximum closer than 30\,g/cm$^{2}$
to the Tunka-133 reconstruction.

The remaining events undergo a full detector simulation and afterwards the same reconstruction algorithm and cuts are used as for the real data.
20 iron and 28 proton events pass all requirements and are used for comparing measurements and simulations.
Since every event contains multiple stations, there are 83 and 124 amplitude measurements for iron and proton, respectively.
\begin{figure}[tb]
	\center
	\includegraphics[width=0.7\textwidth]{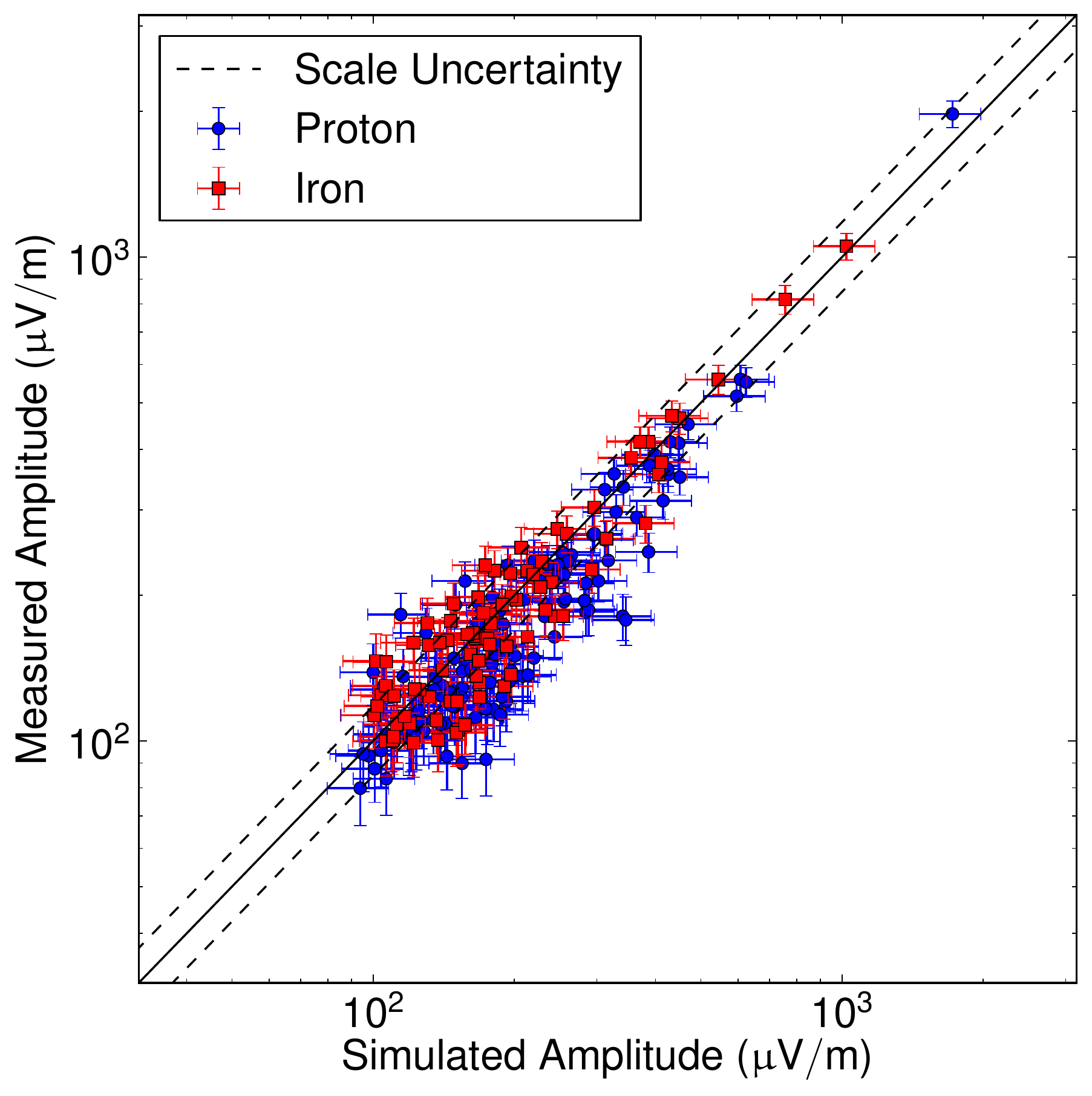}
	\caption{Amplitudes simulated with CoREAS versus measurements from actual showers. 
	For the simulation initial parameters from Tunka-133 where used.
	Error bars indicate the energy ucertainty of Tunka-133 for the simulation and the uncertainty from background for the measurement.}
	\label{fig:coreas}
\end{figure}

In Fig.~\ref{fig:coreas} these ampltiudes are shown versus the measured ones. 
The scale for iron and proton are off by a factor of 0.96 and 0.84 respectively,
within the scale uncertainty of the calibration of 18\%.
Event-to-event and shower-to-shower uncertainties average out for the mean over several antennas from different events.

\section{Summary}
Tunka-Rex is a digital radio array for the detection of air showers in Siberia.
Besides determining the precision and overall benefit of a radio detector in combination with an air-Cherenkov detector,
its goal is to determine the absolute scale of radio emission.
Therefore, we performed an absolute calibration with a commercially available reference source, 
reaching an accuracy for an individual amplitude measurement of 22\%, including 18\% uncertainty on the absolute scale.
The same reference source was also used by LOPES and LOFAR. 
These three experiments now have a consistent amplitude calibration.
Comparing the measured amplitudes to the predictions of the simulation code CoREAS,
we find agreement within the uncertainties.
Therefore, Tunka-Rex confirms the applicability of CoREAS for the simulation of absolute amplitudes of radio emission by air showers.
It also provides a measure of the absolute scale of radio amplitudes in its frequency band.
\clearpage
\acknowledgments{
Tunka-Rex has been funded by the German Helmholtz association (grant HRJRG-303) and 
supported by the Helmholtz Alliance for Astroparticle Physics (HAP).
This work was supported by the Russian Federation Ministry of Education and Science
(agreement 14.B25.31.0010, zadanie 3.889.2014/K) and
the Russian Foundation for Basic Research 
(Grants 12-02-91323, 13-02-00214, 13-02-12095, 14002-10002).}


\begin{thebibliography}{1}
	\bibitem{geomag} F.D.~Kahn and I.~Lerche, \emph{Radiation from cosmic ray air showers}, \emph{Proc. Phys. Soc., Sect. A} {\bf289} 206 (1966).
	\bibitem{askaryan} G.A.~Askaryan, \emph{Excess negative charge of an electron-photon shower and its coherent radio emission}, \emph{Sov. Phys. JETP} {\bf 14} 441 (1961).
	\bibitem{codalemaaskaryan} A.~Belletoile, R.~Dallier, A.~Lecacheux {\it et al.} (CODALEMA Coll.), \emph{Evidence for the charge-excess contribution in air shower radio emission observed by the CODALEMA experiment}, \emph{Astroparticle Physics} {\bf 69} 50-60 (2015).
	\bibitem{lofaraskaryan} P.~Schellart, S.~Buitink, A.~Corstanje {\it et al.} (LOFAR Collaboration), \emph{Polarized radio emission from extensive air showers measured with LOFAR}, \emph{JCAP} {\bf 1410} 014 (2014).
	\bibitem{aeraaskaryan} A.~Aab, P.~Abreu, M.~Aglietta {\it et al.} (Pierre Auger Coll.), \emph{Probing the radio emission from air showers with polarization measurements}, \emph{Phys. Rev. D} {\bf 89} 052002 (2014).
	\bibitem{codalema} D.~Ardouin, A.~Belletoile, D.~Charrier {\it et al.} (CODALEMA Coll.), \emph{Radio-detection signature of high-energy cosmic rays by the CODALEMA experiment}, \emph{NIM A} {\bf 555} 148-163 (2005).
	\bibitem{lopes} H.~{Falcke}, W.D.~Apel, A.F.~Badea {\it et al.} (LOPES Coll.), \emph{Detection and imaging of atmospheric radio flashes from cosmic ray air showers}, \emph{Nature} {\bf 435} 313-316 (2005).
	\bibitem{t133} V.V.~Prosin, S.F.~Berezhnev, N.M.~Budnev {\it et al.} (Tunka-133 Coll.), \emph{Tunka-133: Results of 3 year operation}, \emph{NIM A} {\bf 756} 94-101 (2014).
	\bibitem{taiga} N.M.~Budnev, I.I.~Astapov, A.G.~Bogdanov {\it et al.} (TAIGA Coll.), \emph{TAIGA the Tunka Advanced Instrument for cosmic ray physics and Gamma Astronomy}, \emph{JINST} {\bf 9} C09021 (2014).
	\bibitem{AERAantenna} P.~Abreu, M.~Aglietta, M.~Ahlers {\it et al.} (Pierre Auger Coll.), \emph{Antennas for the Detection of Radio Emission Pulses from Cosmic-Ray induced Air Showers at the Pierre Auger Observatory}, \emph{JINST} {\bf 7} P10011 (2012).
	\bibitem{lopescalib} S.~Nehls, A.~Hakenjos, M.J.~Arts {\it et al.}, \emph{Amplitude calibration of a digital radio antenna array for measuring cosmic ray air showers}, \emph{NIM A} {\bf 589} 350-361 (2008).
	\bibitem{lofarcalib} A.~Nelles, T.~Karskens, M.~Krause {\it et al.}, \emph{Calibrating the absolute amplitude scale for air showers measured at LOFAR}, \emph{JINST} 2015 in preparation.
	\bibitem{lopescalibICRC} K. Link {\it et al.} (LOPES Collaboration), \emph{this issue} 311 (2015).
	\bibitem{trexHWICRC13} R.~Hiller, N.M.~Budnev, O.A.~Gress {\it et al.} (Tunka-Rex Coll.), \emph{The Tunka-Rex antenna station}, In Proceedings of the \emph{33rd ICRC} {\bf 1278} (2013).
	\bibitem{TRexcalib} P.A. Bezyazeekov , N.M. Budnev , O.A. Gress {\it et al.} (Tunka-Rex Collaboration), \emph{Measurement of cosmic-ray air showers with the Tunka Radio Extension (Tunka-Rex)}, \emph{NIM A} (2015) submitted.
	\bibitem{CoREAS} T.~Huege, M.~Ludwig and C.W.~James, \emph{Simulating radio emission from air showers with {CoREAS}}, \emph{AIP Conference Proceedings} {\bf 1535} 128-132 (2013).
	
\end{thebibliography}
\end{document}